\documentclass[twocolumn,showpacs,preprintnumbers,amsmath,amssymb,prl,superscriptaddress]{revtex4}

\usepackage{epsfig}
\usepackage{graphicx}
\usepackage{dcolumn}
\usepackage{bm}

\begin{document}

\title{
Magnon dispersion in the field--induced magnetically ordered phase 
of TlCuCl$_3$}

\author{Masashige Matsumoto}
\altaffiliation{Department of Physics, Faculty of Science, Shizuoka 
University, 836 Oya, Shizuoka 422--8529, Japan}
\affiliation{Theoretische Physik, ETH--H\"onggerberg, CH--8093 Z\"urich, 
Switzerland}

\author{B. Normand}
\affiliation{D\'epartement de Physique, Universit\'e de Fribourg, 
CH--1700 Fribourg, Switzerland}

\author{T. M. Rice}
\affiliation{Theoretische Physik, ETH--H\"onggerberg, CH--8093 Z\"urich, 
Switzerland}

\author{Manfred Sigrist}
\affiliation{Theoretische Physik, ETH--H\"onggerberg, CH--8093 Z\"urich, 
Switzerland}

\date{May 2, 2002}

\begin{abstract}

The magnetic properties of the interacting dimer system TlCuCl$_3$ are 
investigated within a bond--operator formulation. The observed field--induced 
staggered magnetic order perpendicular to the field is described as a Bose 
condensation of magnons which are linear combinations of dimer singlet and 
triplet modes. This technique accounts for the magnetization curve and for 
the field dependence of the magnon dispersion curves observed by high--field 
neutron scattering measurements.

\end{abstract}


\maketitle


\newcommand{\br}{{\mbox{\boldmath$r$}}}
\newcommand{\bk}{{\mbox{\boldmath$k$}}}
\newcommand{\sk}{{\mbox{\footnotesize $k$}}}
\newcommand{\bsk}{{\mbox{\footnotesize \boldmath$k$}}}
\newcommand{\bsQ}{{\mbox{\footnotesize \boldmath$Q$}}}
\newcommand{\bsr}{{\mbox{\footnotesize \boldmath$r$}}}
\newcommand{\bS}{{\mbox{\boldmath$S$}}}
\newcommand{\bQ}{{\mbox{\boldmath$Q$}}}
\newcommand{\bd}{{\mbox{\boldmath$d$}}}
\newcommand{\bsd}{{\mbox{\footnotesize{\boldmath$d$}}}}
\newcommand{\bsigma}{{\mbox{\boldmath$\sigma$}}}
\newcommand{\Tl}{{TlCuCl$_3$}}
\newcommand{\K}{{KCuCl$_3$}}
\newcommand{\ha}{{\hat{a}}}
\newcommand{\hb}{{\hat{b}}}
\newcommand{\hc}{{\hat{c}}}


\Tl~is an insulating quantum spin system with a gap in the spin excitation 
spectrum \cite{Takatsu} at zero field which originates from dimerization 
of the $S=1/2$ spins of the Cu$^{2+}$ ions. The compound is isostructural 
to \K~and the crystal structure can be considered as coupled two--leg ladders
separated by Tl$^+$ ions. However, inelastic neutron scattering (INS) 
measurements of triplet magnon excitations found that the magnon modes have
significant dispersion in all three spatial dimensions for both \Tl~and 
\K~\cite{Kato,Cavadini-1999,Oosawa-2002}. This indicates that these compounds 
are three--dimensional (3d) interacting dimer systems with
interladder interactions stronger than the interdimer interactions within 
the ladders.
As observed in INS experiments by Cavadini {\it et al.},
the magnon modes are split into three by a magnetic field, with splitting
proportional to the field, and the lowest mode becomes soft at a critical field $H_c$
\cite{Cavadini-2002}.
For \Tl~(\K), the zero--field excitation gap and $H_c$
are respectively 0.7meV (2.6meV) and 5.7T (20T).

For $H > H_c$, both compounds show a uniform magnetization parallel to 
the field \cite{Shiramura-1997}. Elastic neutron scattering measurements 
on \Tl~show in addition a staggered magnetic order perpendicular to the 
field \cite{Tanaka}. A Goldstone mode is then expected due to the breaking 
of rotational symmetry around the field axis. Very recently, R\"{u}egg 
{\it et al.} observed such a gapless mode for $H > H_c$ in 
\Tl~\cite{Rueegg-2002-2}, and also reported a renormalized field dependence of 
the higher magnon modes \cite{Rueegg-2002-1}.

Field--induced magnetic order in otherwise gapped ladder systems has been 
described \cite{Tachiki,Mila} by theoretical approaches which focus only 
on the singlet and the lowest triplet magnon. Giamarchi 
and Tsvelik \cite{Giamarchi} cast their theory as a Bose 
condensation of a soft mode, and for \Tl~the Bose condensation of magnons 
has been used \cite{Oosawa-1999,Nikuni} to account for the observed 
temperature dependence of the magnetization. A quantum Monte Carlo simulation
on a simplified 3d cubic lattice was also in agreement with an 
interpretation as magnon condensation \cite{Wessel}.

We note that the structure of the observed staggered order is related to the 
wavevector of the soft magnon \cite{Cavadini-2002,Tanaka}, in 
support of the idea of a magnon Bose condensation. In this paper
we develop a microscopic theory of field--induced magnetic order which 
takes into account the higher triplet modes. These modes, as we will show 
below, must be included to obtain a complete description of the condensate
and of the evolution of the magnon dispersion in the presence of 
field--induced magnetic order. We use a bond--operator formulation which 
retains all four states of each dimer, analogous to the treatment 
of bilayer antiferromagnets by Sommer {\it et al.} \cite{Sommer}.


For the parameterization of the couplings between the Cu$^{2+}$ ion spins,
we follow Ref.~\cite{Cavadini-1999}. The unit cell contains two equivalent 
dimers, and the Hamiltonian is given by
\begin{eqnarray}
{\cal H} &=& \sum_j \left[ \mbox{$J$} \bS_{l,j}^1 \cdot \bS_{r,j}^1
   + \mbox{$J_2'$} \bS_{l,j}^1 \cdot \bS_{r,j+\bsd_2}^1 \right. \label{Hamil} 
\nonumber \\
  &+& \mbox{$J_1$} \sum_{n=l,r} \bS_{n,j}^1 \cdot \bS_{n,j+\bsd_1}^1
   +               \mbox{$J_1'$} \bS_{l,j}^1 \cdot \bS_{r,j+\bsd_1}^1
\nonumber \\
  &+& \mbox{$J_3$} ( \bS_{l,j}^1 \cdot \bS_{l,j+\bsd_{3+}}^2
   +        \bS_{r,j}^1 \cdot \bS_{r,j+\bsd_{3-}}^2 ) \\
  &+& \left. \mbox{$J_3'$} ( \bS_{l,j}^1 \cdot \bS_{r,j+\bsd_{3+}}^2
       +    \bS_{r,j}^1 \cdot \bS_{l,j+\bsd_{3-}}^2 ) \right]
       + \mbox{$[ 1 \leftrightarrow 2 ]$}. \nonumber 
\end{eqnarray}
Here $\bS_{n,j}^m$ is the spin $S=1/2$ operator in unit cell $j$ on the 
sublattice $m=1,2$, and $n$( $=l,r$) denotes the left or right spin of the 
dimer. $\bd_1=\ha$, $\bd_2=2\ha+\hc$, and $\bd_{3\pm}=\hb/2\pm(\ha+\hc/2)$,
where $\ha$, $\hb$, and $\hc$ are unit vectors corresponding to the $a$, 
$b$, and $c$ axes, respectively \cite{Cavadini-1999}. 

Because the intradimer exchange coupling $J$ is the largest
\cite{Cavadini-1999,Oosawa-2002}, we introduce bond operators $s^\dagger$ 
and $t_\alpha^\dagger$ based on these dimers \cite{Sachdev}. In the presence 
of an external field, the appropriate operators are \cite{Normand-2000}
\begin{eqnarray}
s^\dagger |0\rangle & = & (|\uparrow \downarrow \rangle - |\downarrow 
\uparrow \rangle)/\sqrt{2}, \;\; t_+^\dagger |0\rangle = -|\uparrow 
\uparrow \rangle, \nonumber \\
t_0^\dagger |0\rangle & = & (|\uparrow \downarrow \rangle + |\downarrow 
\uparrow\rangle)/\sqrt{2}, \;\; t_-^\dagger |0\rangle = |\downarrow 
\downarrow \rangle. \nonumber 
\end{eqnarray}
These have Bose statistics and are subject to the constraint $s^\dagger s 
+ \sum_\alpha t_\alpha^\dagger t_\alpha = 1$ at each dimer $\{j,m\}$, where 
$\alpha = +,0,-$. We introduce the Fourier transformation 
$c_j^{m\dagger} = (1/N) \sum_\bsk c_\bsk^{m\dagger} e^{i\bsk\cdot\bsr_j^m}$ 
for the dimer operators, where $2N$ is the total dimer number.

\begin{figure}[t!]
\epsfxsize=6.9cm
\epsfbox{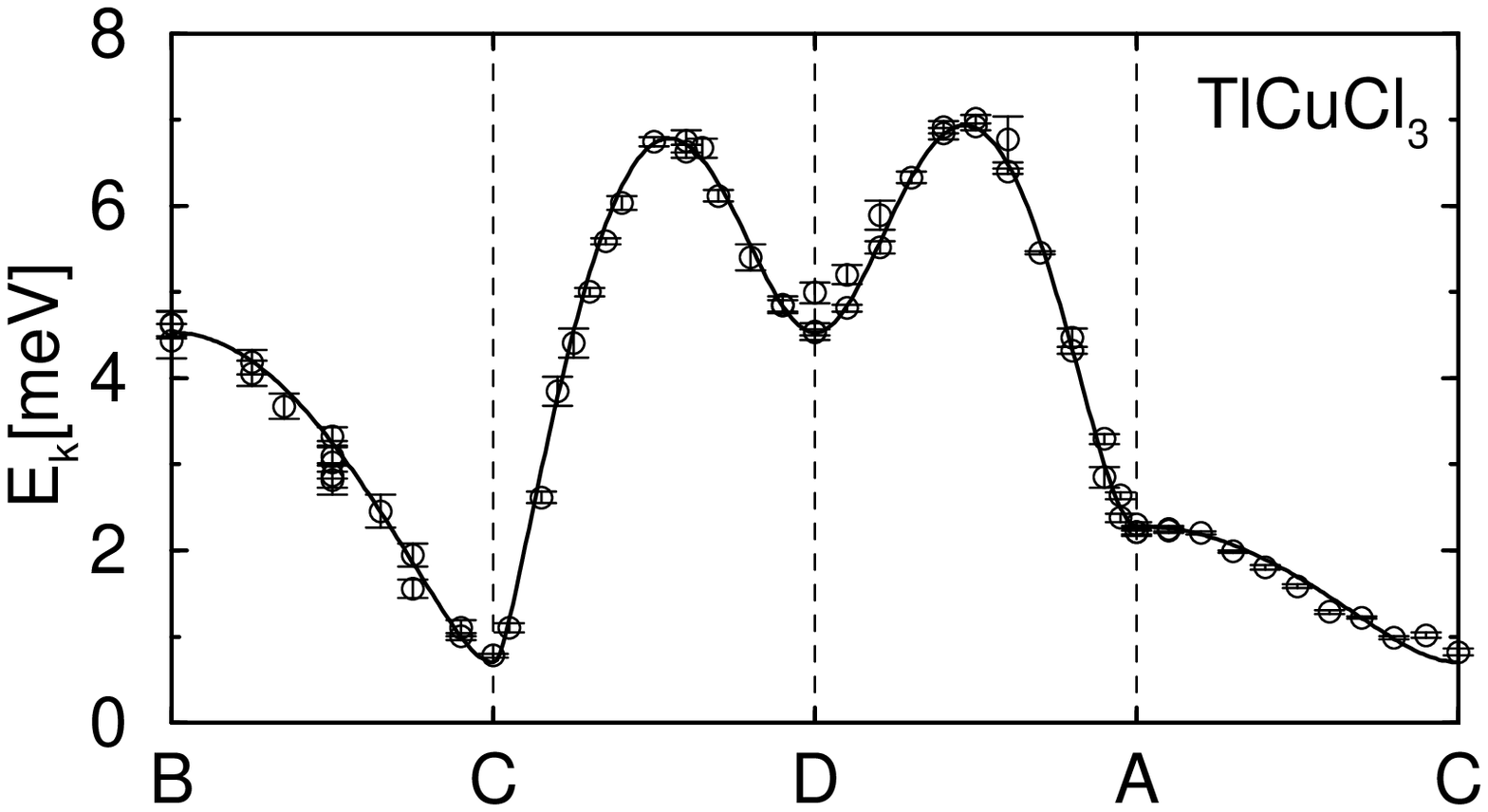}
\epsfxsize=6.9cm
\epsfbox{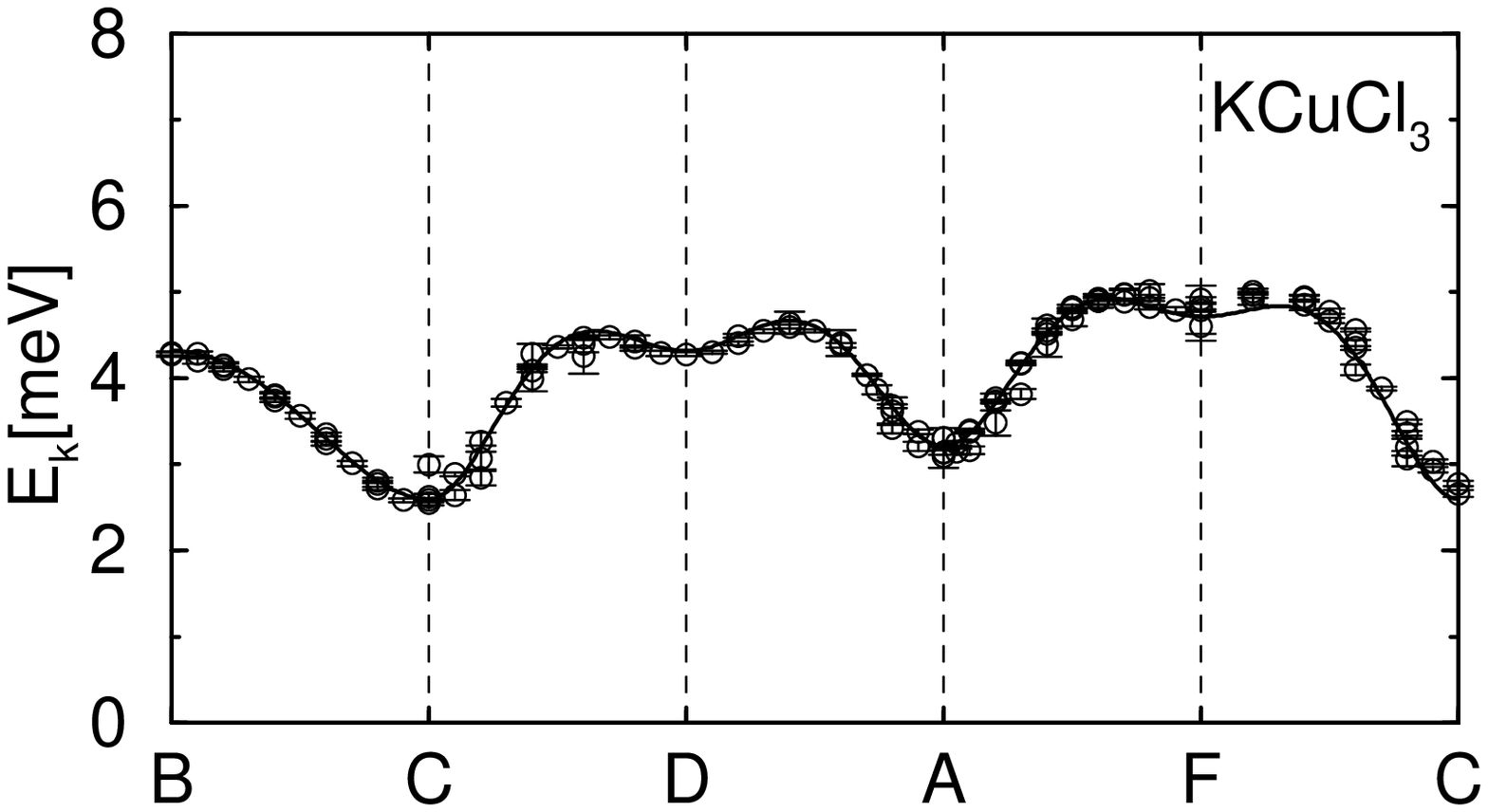}
\caption{
Zero--field magnon dispersion relations for \Tl~and \K. The $x$--axis labels 
represent the reciprocal--space points B=(0,2$\pi$,2$\pi$), C=(0,0,2$\pi$), 
D=(0,0,0), A=($\pi$,0,0), and F=($\pi$,0,2$\pi$). Experimental results 
\protect\cite{Cavadini-1999} are shown as points, and were measured at 
$T=1.5$K (5K) for \Tl~(\K). Solid lines are theoretical results using the 
parameters listed in Table I.
}
\label{fig:1}
\end{figure}

We restrict our considerations to zero temperature, and begin with the 
low--field region ($H \le H_c$). Here the dimer singlets have the lowest 
energy, so the $s$--bosons are taken to be condensed. The $s$-operators 
are replaced by a $c$--number $s_\bsk^m = \bar{s}_0 \delta_{\bsk,0}$ 
\cite{Gopalan}, and the local constraint relaxed to the global one 
${\bar{s}_0}^* \bar{s}_0 = N - \sum_{\bsk,\alpha} {t_{\bsk\alpha}^m}^\dagger 
t_{\bsk\alpha}^m$. Retaining quadratic terms in the triplet operators, 
\begin{eqnarray}
&& {\cal H} = \sum_{m,\bsk} \{
       \sum_\alpha [ (J - \alpha g \mu_{\rm B}H + f_\bsk)
             t_{\bsk\alpha}^{m\dagger} t_{\bsk\alpha}^m
           + g_\bsk t_{\bsk\alpha}^{m\dagger} t_{\bsk\alpha}^{\bar m}] \cr
&& ~~~~~~~~~~~~ + \sum_{\alpha=+,-}
         ( f_\bsk t_{\bsk\alpha}^m t_{-\bsk\bar{\alpha}}^m
         + g_\bsk t_{\bsk\alpha}^m t_{-\bsk\bar{\alpha}}^{\bar m} + 
         {\rm H. c.} ) \cr
&& ~~~~~~~~~~~~ + ( f_\bsk t_{\bsk 0}^m t_{-\bsk 0}^m
            + g_\bsk t_{\bsk 0}^m t_{-\bsk 0}^{\bar m} + {\rm H. c.} )/2 \},
\label{eqn:Hlow}
\end{eqnarray}
where
$f_\bsk = J_a \cos k_x + J_{a2c} \cos(2k_x+k_z)$,
$g_\bsk = 2 J_{abc} \cos(k_x + k_z/2) \cos(k_y/2)$,
$\bar{\alpha}=-\alpha$, and $\bar{m}=2,1$ for $m=1,2$.
The effective interdimer interactions \cite{Cavadini-1999} are 
given by $J_a=J_1-J_1'/2$, $J_{a2c}=-J_2'/2$, and $J_{abc}=(J_3-J_3')/2$,
where we note changes in the signs of the parameters $J_{a2c}$ and,
for $J_1' > 2 J_1$, $J_a$.
Introducing the operators $t_{\bsk\alpha}^\pm = (t_{\bsk\alpha}^1 
\pm t_{\bsk\alpha}^2) / 
\sqrt{2}$ gives two independent ($\pm$) modes \cite{Normand-1996} whose 
eigenvalues are obtained by the Bogoliubov transformation \cite{Normand-2000}
\begin{equation}
  E_{\bsk\alpha}^\pm = \sqrt{(J +f_\bsk\pm g_\bsk)^2-(f_\bsk\pm g_\bsk)^2}
  -\alpha g \mu_{\rm B}H.
\label{eqn:disp}
\end{equation}
The Brillouin zone lies between $-\pi$ and $\pi$ (unit lattice spacing)
in each direction, and the two branches correspond to the two--sublattice 
system. We treat only the $+$ mode in the expanded Brillouin zone
($-2\pi\le k_z \le 2\pi$) in the $z$--direction, because the magnon 
dispersions obey the relations $E_{\bsk\alpha}^- = E_{\bsk\pm(0,0,2\pi), 
\alpha}^+$. We extract the effective interactions ($J$, $J_a$, $J_{a2c}$, 
$J_{abc}$) from the data at zero field for both \Tl~and 
\K~(Fig. \ref{fig:1}), and these are listed in Table I. The values are 
consistent with the results of Ref.~\cite{Oosawa-2002}. They are also 
similar to those of Ref.~\cite{Cavadini-1999}, but not identical because 
the current treatment [leading to (\ref{eqn:disp})] goes beyond a pure 
dimer description.
The signs are consistent with the expectation 
that all intersite parameters $\{J\}$ in (1) are antiferromagnetic.

\begin{figure}[t!]
\epsfxsize=7cm
\epsfbox{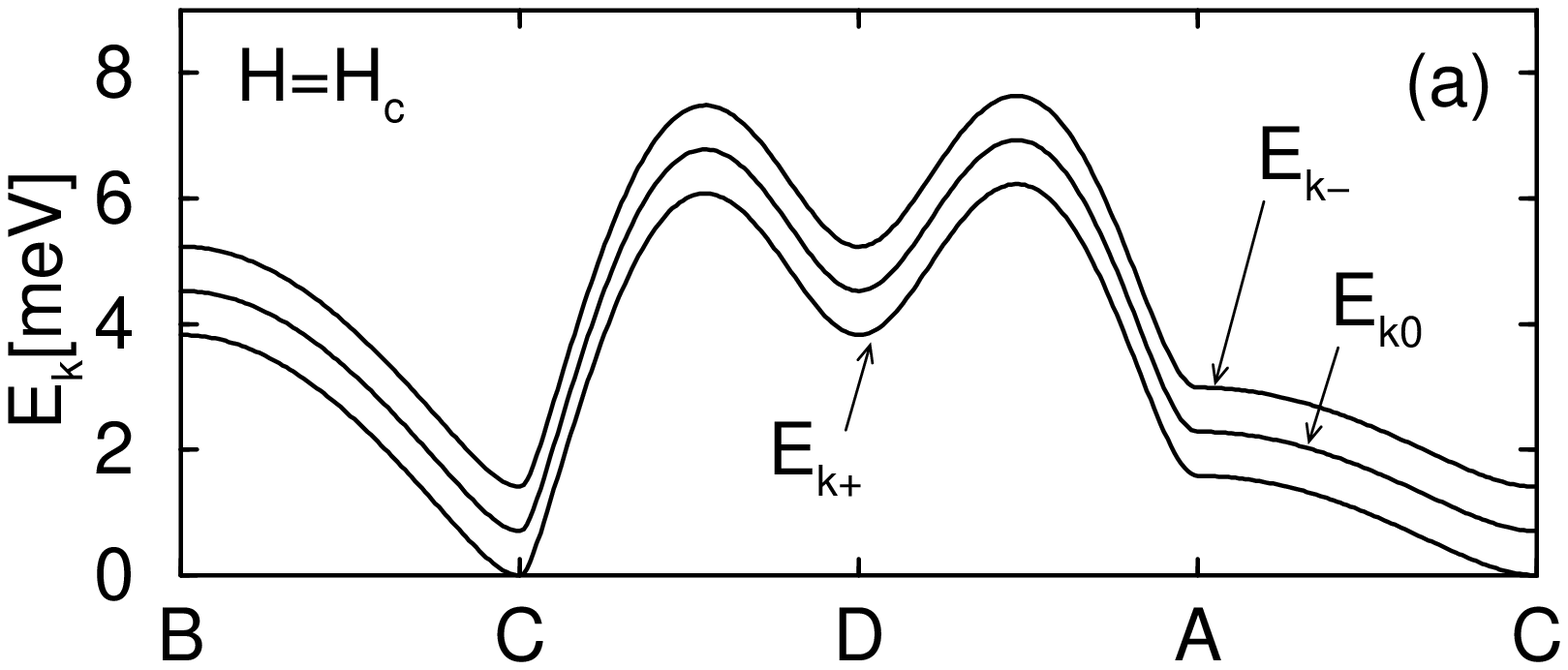}
\epsfxsize=7cm
\epsfbox{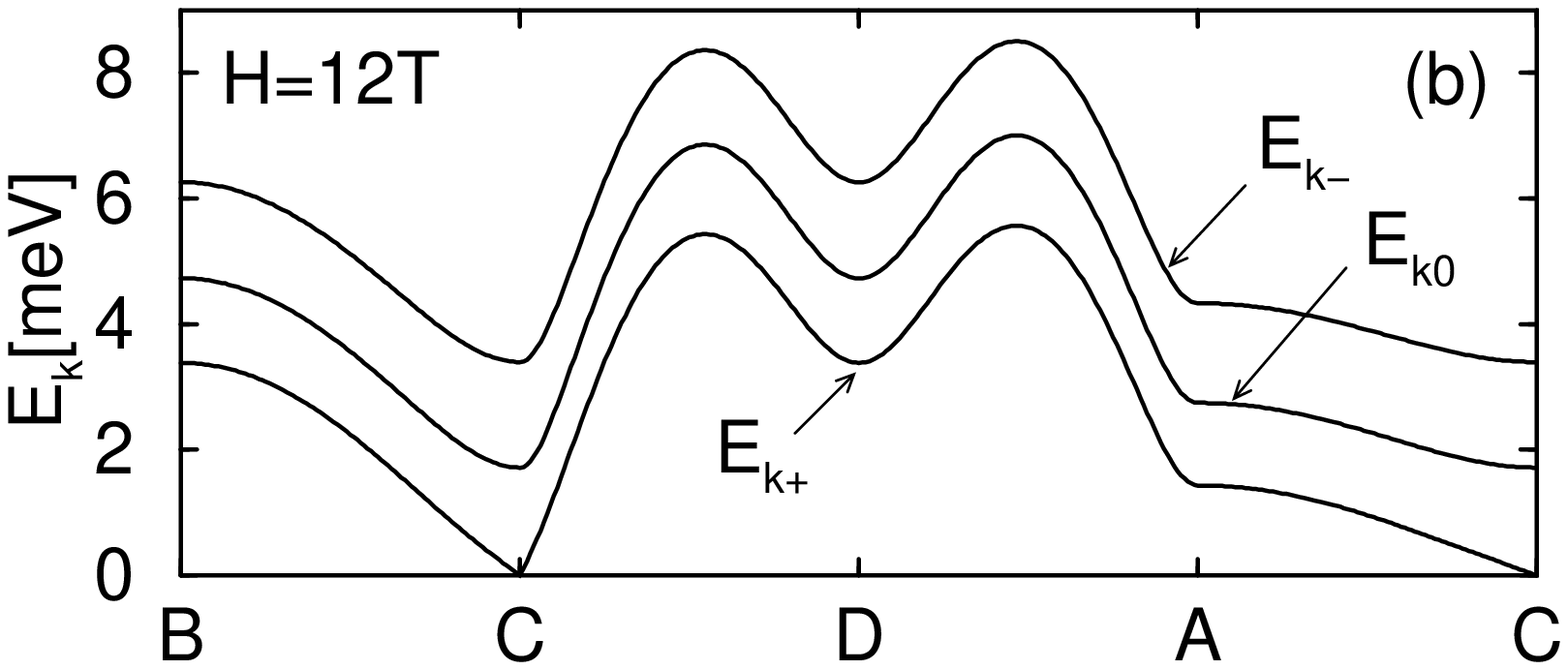}
\caption{
Calculated magnon dispersions for \Tl~at $H=H_c(=5.6$T$)$ (a), and $H>H_c$ 
($H = 12$T) (b).
}
\label{fig:2}
\end{figure}

The three magnon modes are degenerate in zero field, and, as shown in 
Fig.~\ref{fig:2}, are split linearly by an external field in agreement 
with INS results \cite{Cavadini-2002}. Below $H_c$, a $t_+^\dagger$ 
triplet excited from the singlet condensate may propagate due to the 
interaction between triplets and singlets. The wave function of this 
excited state can be approximated by a linear combination of dimer singlets 
and triplets as 
\begin{equation}
|\psi\rangle \sim u (|\uparrow\downarrow\rangle - |\downarrow\uparrow\rangle)
  - v e^{i (\bsk \cdot\bsr_j^m - E_{\bsk +}t)} |\uparrow\uparrow\rangle,
\label{eqn:psi}
\end{equation}
where $u$ is of order unity and $v$ is a small, real coefficient. The 
expectation values of the spin operator components at a given dimer 
$j,m$ in the state (\ref{eqn:psi}) are
\begin{eqnarray}
&& \langle S_{l,z} \rangle_{j,m} = \langle S_{r,z} \rangle_{j,m} \sim v^2/2, 
\\
&& \langle S_{l,x} \rangle_{j,m} = -\langle S_{r,x} \rangle_{j,m}
         \sim (u v / 2) \cos(\bk\cdot\br_j^m - E_{\bsk +}t), \nonumber \\
&& \langle S_{l,y} \rangle_{j,m} = -\langle S_{r,y} \rangle_{j,m}
         \sim - (u v / 2) \sin(\bk\cdot\br_j^m - E_{\bsk +}t). \nonumber 
\label{eqn:spin}
\end{eqnarray}
This excited mode has a very small, uniform magnetic moment parallel to 
the field, and thus gains Zeeman energy. Perpendicular to the field,
it also possesses a finite magnetic moment, which is characterized by the 
wave vector $\bk$ and energy $E_{\bsk,+}^+$, and is staggered (spins $l$ 
and $r$ oppositely aligned) due to the antiferromagnetic intradimer coupling.
For the mode $t_-^\dagger$, the direction of the uniform magnetic moment is 
antiparallel to the field, leading to a higher Zeeman energy. Finally, the 
$t_0^\dagger$ mode has no magnetization perpendicular to the field, and a 
moment parallel to the field which is modulated with wavevector $\bk$,
so its energy does not shift with the field. On increasing the field,
these modes shift position without changing the shape of their dispersion 
(\ref{eqn:disp}), and the lowest ($\alpha=+$) mode becomes gapless at the
point C with $\bQ\equiv (0,0,2\pi)$ [Fig.~\ref{fig:2}(a)], which determines 
the critical field. Thus at $H=H_c$, the lowest mode exhibits a quadratic 
dependence on $\bk$ around $\bQ$.

\begin{table}[b!]
\caption{Effective dimer interactions.}
\begin{ruledtabular}
  \begin{tabular}{l|r|r}
    \multicolumn{1}{r|}{}   &   \Tl   &   \K    \\ \hline
    $J$~~~~~[meV]             &   5.501 &   4.221 \\
    $J_a$~~~~[meV]            &  -0.215 &  -0.212 \\
    $J_{a2c}$~~[meV]            &  -1.581 &  -0.395 \\
    $J_{abc}$~~[meV]            &   0.455 &   0.352 \\
  \end{tabular}
\end{ruledtabular}
\end{table}

To describe the field regime with $H > H_c$, the Hamiltonian 
(\ref{eqn:Hlow}) must be extended to include triplet--triplet interactions.
The coefficients of these terms involve combinations of the intersite 
exchange constants in (1) beyond the interdimer interactions in 
(\ref{eqn:Hlow}). To determine these we have made the simplifying assumption 
that $J_1=J_3'=0$, so that the three remaining coefficients, $J_1'$, $J_2'$, 
and $J_3$, are specified by the interdimer interactions. Because the 
additional triplet--triplet interactions are largely governed by terms 
involving $J_2'$, this is not a significant approximation. A further 
assumption is made by neglecting terms involving three $t$ operators, 
which give only small corrections concentrated in the region of maximum 
staggered magnetic order.
For $H>H_c$, Bose condensation of the lowest triplet
implies a macroscopic occupation of the $t_{\bsk +}$ mode at $\bQ$.

For a full description of this regime,
the singlet and triplet operators are transformed \cite{Sommer} to 
\begin{eqnarray}
&& a_\bsk^m = u s_\bsk^m
            + v ( x t_{\bsk+\bsQ,+}^m + y t_{\bsk+\bsQ,-}^m ), \nonumber \\
&& b_{\bsk +}^m = u ( x t_{\bsk +}^m + y t_{\bsk -}^m )
                - v s_{\bsk+\bsQ}^m, \\
&& b_{\bsk 0}^m = t_{\bsk 0}^m,~~~~~~b_{\bsk -}^m = x t_{\bsk -}^m - y 
t_{\bsk +}^m. \nonumber
\label{eqn:trans}
\end{eqnarray}
The $\bk$--independent coefficients $u$, $v$, $x$, and $y$ arise from 
two unitary transformations, and may be written as $u=\cos\theta$, 
$v=\sin\theta$, $x=\cos\phi$, $y=\sin\phi$, with $\theta$ and $\phi$
to be determined. We treat the $a_\bsk^m$ operator as uniformly 
condensed, $a_\bsk^m=\bar{a}_0 \delta_{\bsk,0}$, and the ground state 
as a coherent condensate of the $a_0^m$ operator. We emphasize that the 
highest triplet mode ($t_{\bsQ -}^m$) also participates in the condensate,
because $ {\cal H} $ contains processes $t_+^\dagger t_-^\dagger s s$ 
which nucleate $t_+$ and $t_-$ triplets from singlets. The linear 
combination of singlet and triplets in the condensate $a_0^m$ yields a 
staggered magnetization perpendicular to the field with wave vector 
$\bQ$, as observed in Ref.~\cite{Tanaka}.

With the transformation of Eq.~(\ref{eqn:trans}), the Hamiltonian to 
quadratic order in the $b$ operators takes the form $ {\cal H} = O(b^0) + 
O(b^1) + O(b^2)$. Because the particle number is unaltered, the $c$-number 
$\bar{a}_0$ may be replaced by ${\bar{a}_0}^* \bar{a}_0 = N - 
\sum_{\bsk,\alpha} {b_{\bsk\alpha}^m}^\dagger b_{\bsk\alpha}^m$.
The constant terms ($O(b^0)$) represent the mean--field energy of the 
$a_0$--condensate, and the parameters ($\theta$, $\phi$) are chosen 
to minimize this energy, which also eliminates the $O(b^1)$ terms in the 
transformed Hamiltonian. The critical field $H_c$ is determined by the 
condition $\theta\rightarrow 0$, which gives purely singlet character to 
the condensate. The limit ($\theta=\pi/2$, $\phi=0$) gives 
a condensate with purely triplet $t_+$ character, and determines the 
saturation field $H_s$ for full parallel polarization.
The values for $H_c$ and $H_s$ are
\begin{eqnarray}
H_c &=& \sqrt{J^2-2J(|J_a|+|J_{a2c}|+2J_{abc})}/(g \mu_{\rm B}), \cr
H_s &=& (J+2|J_a|+2|J_{a2c}|+4J_{abc})/(g \mu_{\rm B}),
\label{eqn:HC}
\end{eqnarray}
where the form of $H_c$ coincides with the soft--mode condition in the 
low--field regime. 

The $O(b^2)$ terms are diagonalized by a Bogoliubov transformation which 
yields the energies of the collective modes of the condensate. Above $H_c$, 
the lowest mode remains gapless, but develops a linear dependence on $\bk$ 
near $\bQ$ [Fig.~\ref{fig:2}(b)]. This is a Goldstone mode: a staggered 
magnetic moment $M_\bot$, whose mean--field value is $M_\bot = u v (x 
+ y)/\sqrt{2}$, is induced perpendicular to the magnetic field, and 
this breaks rotational symmetry around the axis parallel to the field.
Rotations of this induced staggered moment are realized by changing the 
phase of $x$ and $y$ in Eq.~(\ref{eqn:trans}) ($x\rightarrow e^{-i\chi}x$,
$y\rightarrow e^{i\chi}y$, where $\chi$ is the rotation angle), and do not 
change the energy.
\begin{figure}[t!]
\epsfxsize=7.5cm
\epsfbox{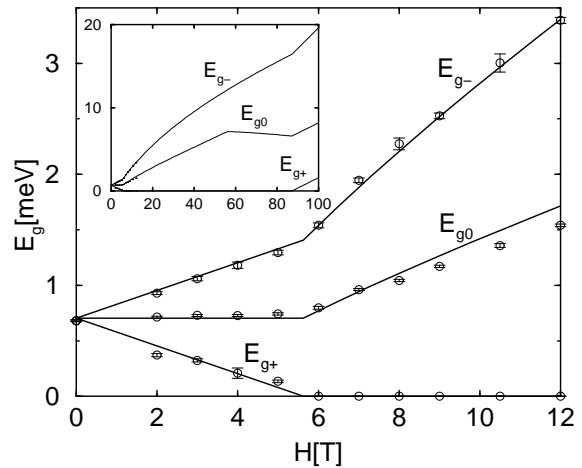}
\caption{
Field dependence of the energy gap of the three magnon modes in \Tl.
Points are data from the INS experiment of Ref.~\protect\cite{Rueegg-2002-1},
and solid lines the results of the present theory.
Inset: field dependence at high fields. We have used $g=2.16$ for the 
$g$--factor of \Tl~\protect\cite{Rueegg-2002-1}.
}
\label{fig:3}
\end{figure}
As a result the Goldstone mode remains gapless, and the
field dependence of the higher modes is also renormalized, in agreement 
with experiment (Fig.~\ref{fig:3}). 
The energy gaps of the higher modes $E_{g0}$ and $E_{g-}$ show an abrupt 
increase in slope at $H_c$ (Fig.~\ref{fig:3}). At the saturation field 
$H_s$, the $\bk$ dependence of the lowest excitation mode near $\bQ$ 
becomes quadratic again.
In the high--field region above $H_s$, the condensate consists only of the 
lowest--lying triplet, and a gap reopens in the spectrum of the lowest 
(pure singlet) excitation mode (inset Fig. \ref{fig:3}).

\begin{figure}[t!]
\epsfxsize=4.2cm
\epsfbox{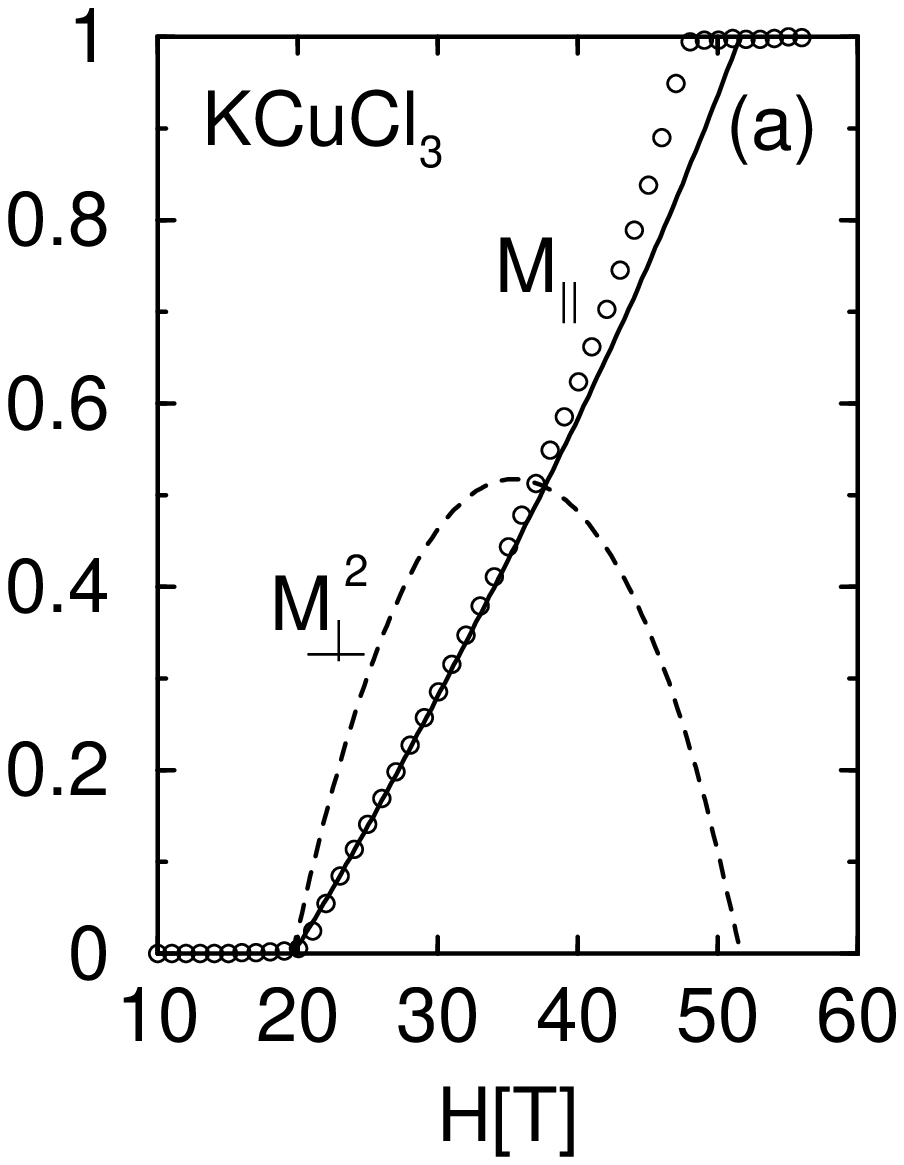}
\epsfxsize=4.3cm
\epsfbox{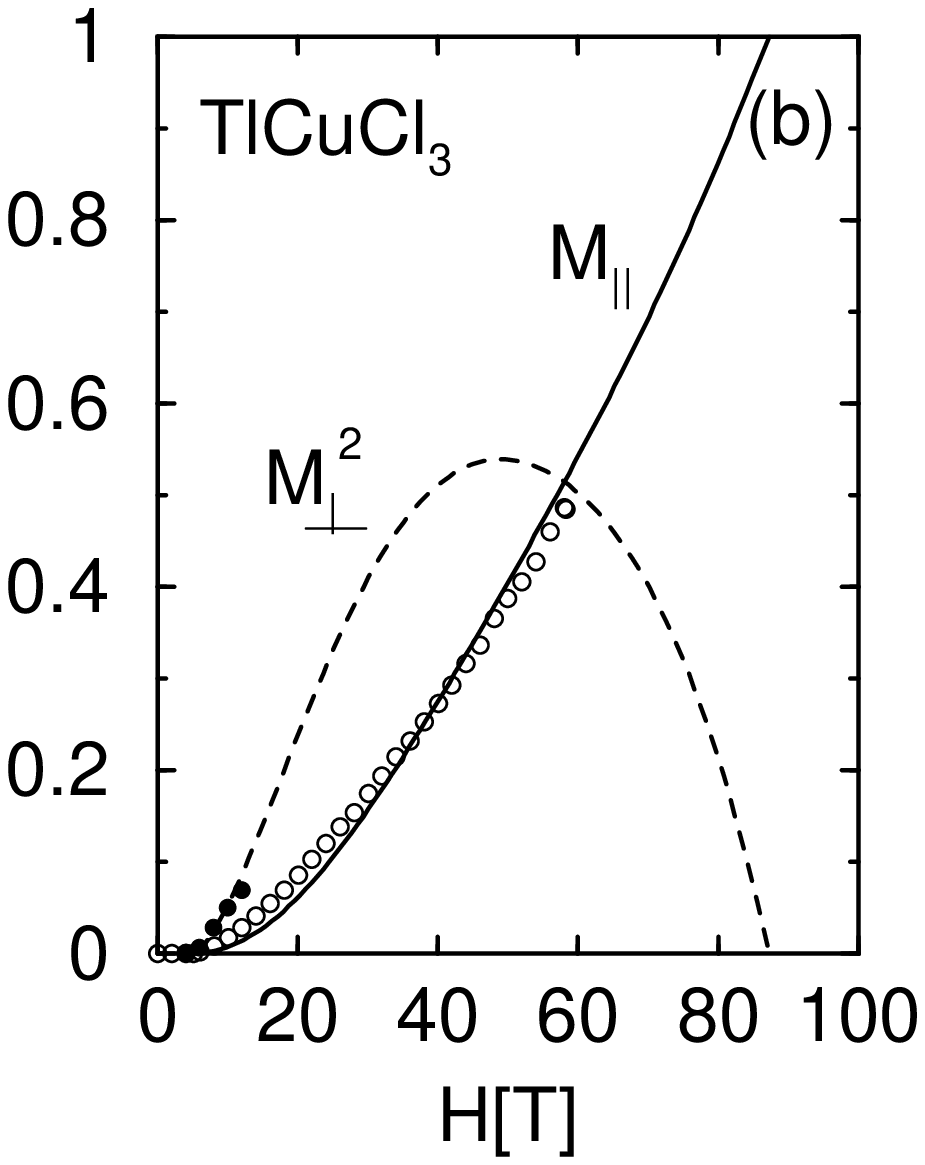}
\caption{
Normalized magnetization curves for \K~(a), and \Tl~(b).
The solid line  ($M_\|$) is the magnetization parallel to the field and the dashed line 
($M_\bot^2$) the square of the magnetization perpendicular to the field.
Open (filled) points are experimental data for $M_\|$ ($M_\bot^2$)
measured at $T=1.3$K (0.2K) \protect\cite{Oosawa-2002-2,Tanaka},
where $g=2.29$ is the $g$--factor for \K. For \Tl~we have used the same 
$g$ as in Fig.~\ref{fig:3}.
}
\label{fig:4}
\end{figure}

We may also consider the parallel and perpendicular magnetization curves.
For \Tl~(\K), the parameters of Table I give a critical field of $H_c=5.6$T 
($19.6$T), which is consistent with the measured value \cite{Shiramura-1997}.
For both \Tl~and \K,~the square of the staggered moment ($M_\bot^2$) has 
linear dependence close to $H_c$ and to $H_s$ (Fig.~\ref{fig:4}), indicating 
that $M_\bot \sim \sqrt{H-H_c}$. For \K,~the magnetization parallel to the 
field ($M_\|$) is almost linear in $H$, as shown in Fig. \ref{fig:4}(a),
in good agreement with experiment. For \Tl,~the field dependence of $M_\|$ 
appears not to be linear in $H$, again in very good agreement with the 
observed form \cite{Shiramura-1997}. The theoretical mean--field value is 
given by $M_\| = v^2 (x^2-y^2)$, from which it is clear that this difference 
is due to the magnitude of the interdimer interactions. If the contribution 
to the $a_0^m$ operator of the highest triplet mode ($t_-$) is neglected
[Eq. (\ref{eqn:trans})], $M_\|$ would become completely linear in $H$.
However, the strong interdimer interactions involve the highest mode largely
through interactions of the type $t_+^\dagger t_-^\dagger s s$, which cost 
Zeeman energy, thus reducing the value of $v$ for the $a_0^m$ operator, and 
consequently $M_\|$ is suppressed near $H_c$ for \Tl.

We have studied the evolution of the magnon dispersion in \Tl~and 
\K~as the magnetic field is tuned through the quantum critical point 
separating a gapped spin--liquid state from a state of field--induced 
staggered magnetic order which exists between the critical and 
saturation fields. A consistent theory requires the 
admixture of both the lowest and highest triplet modes into the singlet 
dimer state to form a Bose condensate. The spectrum has a gapless 
Goldstone mode associated with the breaking of rotational symmetry by 
the staggered magnetic order, as observed by R\"{u}egg {\it et al.} 
\cite{Rueegg-2002-2} for \Tl. The two higher excitation modes are also renormalized, 
in further agreement with observation \cite{Rueegg-2002-1}. Finally, our 
zero--temperature mean--field description is also in good accord with
measurements \cite{Shiramura-1997,Oosawa-2002-2,Tanaka} of the uniform and staggered 
magnetization for both \K~and \Tl.


We express our sincere thanks to N. Cavadini, K. Kindo, H. Kusunose,
A. Oosawa, Ch. R\"{u}egg, and H. Tanaka
for valuable discussions and for provision of experimental data.
This work is supported by the Japan Society for the Promotion of Science
and the Swiss National Fund.

\end{document}